\documentclass[reprint,superscriptaddress,amsmath,amssymb,aps,prx]{revtex4-1}

\usepackage{graphicx}
\usepackage{dcolumn}
\usepackage{bm}
\usepackage{hyperref}
\usepackage{notes2bib}
\usepackage{siunitx}

\newcommand{\beginsupplement}{%
	\setcounter{section}{0}
	\renewcommand{\thesection}{S\arabic{section}}%
	\setcounter{figure}{0}
	\renewcommand{\figurename}{Supplemental Figure}%
}

\begin{document}

\title{Combined minivalley and layer control in twisted double bilayer graphene}

\author{Folkert K. de Vries}
\email{devriesf@phys.ethz.ch}
\affiliation{Laboratory for Solid State Physics, ETH Z\"{u}rich, CH-8093 Z\"{u}rich, Switzerland}

\author{Jihang Zhu}
\affiliation{Department of Physics, The University of Texas at Austin, Austin, Texas 78712, USA}

\author{El\'{i}as Portol\'{e}s}
\author{Giulia Zheng}
\author{Michele Masseroni}
\author{Annika Kurzmann}
\affiliation{Laboratory for Solid State Physics, ETH Z\"{u}rich, CH-8093 Z\"{u}rich, Switzerland}

\author{Takashi Taniguchi}
\author{Kenji Watanabe}
\affiliation{National Institute for Material Science, 1-1 Namiki, Tsukuba 305-0044, Japan}

\author{Allan H. MacDonald}
\affiliation{Department of Physics, The University of Texas at Austin, Austin, Texas 78712, USA}

\author{Klaus Ensslin}
\author{Thomas Ihn}
\author{Peter Rickhaus}
\email{peterri@phys.ethz.ch}
\affiliation{Laboratory for Solid State Physics, ETH Z\"{u}rich, CH-8093 Z\"{u}rich, Switzerland}

\date{\today}

\begin{abstract}
Control over minivalley polarization and interlayer coupling is demonstrated in double bilayer graphene twisted with an angle of 2.37$^\circ$.
This intermediate angle is small enough for the minibands to form and large enough such that the charge carrier gases in the layers can be tuned independently.
Using a dual-gated geometry we identify and control all possible combinations of minivalley polarization via the population of the two bilayers.
An applied displacement field opens a band gap in either of the two bilayers, allowing us to even obtain full minivalley polarization.
In addition, the carriers, formerly separated by their minivalley character, are mixed by tuning through a Lifshitz transition, where the Fermi surface topology changes.
The high degree of control over the minivalley character of the bulk charge transport in twisted double bilayer graphene offers new opportunities for realizing valleytronics devices such as valley valves, filters and logic gates.
\end{abstract}

\maketitle
\begin{figure}[t]
\includegraphics{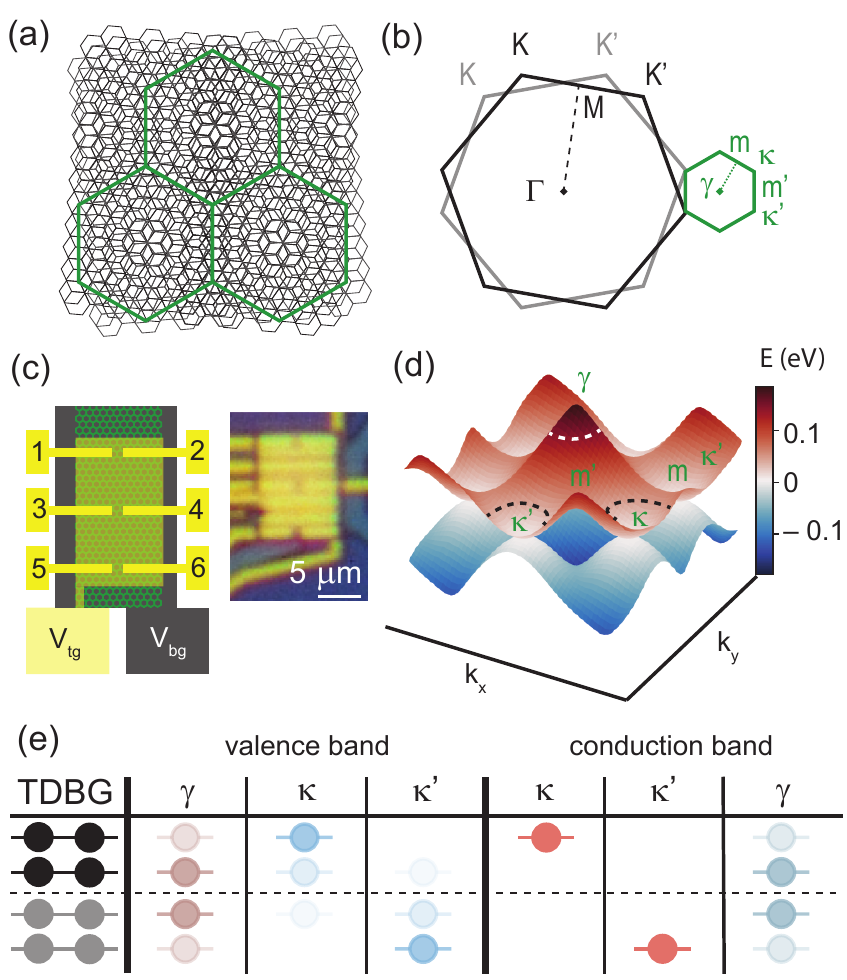}
\caption{\label{fig:1} (a) Schematic representation of twisted double bilayer graphene (TDBG), forming moir\'{e} unit cells indicated by the green hexagons.
(b) Sketch of the Brillouin zone (BZ) of the two bilayers in TDBG (gray and black), leading to the definition of a mini-BZ (green). The characteristic points of the mini-BZs are indicated. 
(c) The left panel depicts the device measured, with TDBG (green), graphite bottom gate $V_\mathrm{bg}$ (dark grey), global top gate $V_\mathrm{tg}$ (opaque yellow) and numbered Ohmic contacts (yellow). The right panel shows an optical image of the device. 
(d) Lowest energy (first moir\'{e}) valence and conduction band of TDBG with a twist of 2.3$^{\circ}$, calculated using a continuum model described in the Supplemental Material~\cite{SM}. The characteristic points of the mini-BZ are indicated and Fermi contours around these points are sketched with a dotted line.
(e) Schematic representation of the calculated wavefunction probability distribution over the four graphene layers for the conduction and valence band at zero displacement field, labeled by the origin of their Fermi surfaces in reciprocal space. The color represents the charge (electrons are red, holes are blue) and the opacity the probability. For details see the Supplemental Material~\cite{SM}.}
\end{figure}

The bandstructure of graphene exhibits two minima at the same energy known as 'valleys'.
These valleys could form the basis for a fundamentally new type of electronics~\cite{Rycerz_2007,Schaibley_2016}.
Experimental realizations of such valleytronic devices include valley polarized transport~\cite{Ju_2015}, the valley Hall effect~\cite{Gorbachev_2014} or a valley valve~\cite{Li_2018}.
Recently, twisting two graphene sheets has opened up a new family of two-dimensional hexagonal systems having a valley degree of freedom as well. 
The twist leads to a superlattice, or moir\'{e} pattern, which introduces a new unit cell in real space [Fig.~\ref{fig:1}(a)] and a new or mini unit cell in reciprocal space with its own minivalleys [Fig.~\ref{fig:1}(b)] that has remarkable effects on the bandstructure~\cite{Bistritzer_2011}.
At small twist angles, below 1$^\circ$, the layers are coupled and the superlattice leads to insulating regions with different topology~\cite{SanJose_2013}, resulting in topological edge channels between these regions~\cite{Rickhaus_2018}.
For angles of $\sim\,1^\circ$, isolated flat bands arise in the coupled layers for twisted bilayer graphene (TBG)~\cite{SuarezMorell_2010} and the bands are flattened in twisted double bilayer graphene (TDBG)~\cite{Lee_2019,Liu_2019_theory}. %Koshino_2019
Flattening of bands boosts many-body interactions and leads to the formation of correlated states such as insulators and superconductors. These states have been measured in TBG~\cite{Cao_2018_insulator,Cao_2018_superconductor}. Recently correlated states were observed in TDBG as well~\cite{Burg_2019,Shen_2019,Liu_2019}.
For large twist angles ($\sim\,10^\circ$ or larger) the layers are decoupled~\cite{Rickhaus_2019_cfield}. The charge carrier wavefunctions are then bound to one of the two layers only, and can therefore be tuned independently using a dual-gated structure~\cite{Rickhaus_2019_cfield,Rickhaus_2019_thickness}. 
While the large and small angle devices have been studied thoroughly, the angles in-between ($\sim\,2^\circ$) have received limited attention in experiments~\cite{Li_2010,Kim_2016_lifshitz,Cao_2016}.
Potential correlated states aside, these intermediate angles are relevant in terms of layer and minivalley tunability since the energy of the cross-over regime between coupled and decoupled wavefunctions can be accessed by gating.

Our TDBG device with an angle of 2.37$^\circ$ [Fig.~\ref{fig:1}(c)] exhibits a bandstructure resembling bilayer graphene but on a much smaller energy scale, as shown in Fig.~\ref{fig:1}(d).
A global bottom gate and top gate enable us to tune the density and displacement field in the two bilayers, creating a tunable testbed to explore the minibands and minivalleys.
The displacement field controls the band gaps in each of the bilayers~\cite{Oostinga_2008,Rickhaus_2019_cfield}.
The different electronic states expected from bandstructure calculations are introduced in Fig.~\ref{fig:1}(d,e), sorted by the characteristic point in the mini-Brillouin zone ($\kappa$, $\kappa'$ or $\gamma$) enclosed by the Fermi surfaces.
The $\kappa$- and $\kappa'$-states are mostly confined to one bilayer each and therefore decoupled.
This decoupling provides control over the layer population via the electrostatic top/bottom gates. 
Since the carriers in the decoupled bilayers are minivalley polarized, the gates directly control the minivalley character of the bulk transport in the device.
Such gate control over the (mini)valley degree of freedom separates TDBG from graphene devices.
Perfect minivalley polarization can be achieved when the Fermi level is tuned into the band gap at $\kappa$ or $\kappa'$ by tuning the displacement field. 
This is in contrast with TBG where only partial minivalley polarization can be reached~\cite{Berdyugin_2019}.
Contrary to $\kappa$- and $\kappa'$-states, carriers stemming from the $\gamma$-points occupy all layers [Fig.~\ref{fig:1}(e)], i.e. the two bilayers are coupled.
Tuning the Fermi level to these $\gamma$-states involves going through a Lifshitz transition, a change in the Fermi surface topology~\cite{Lifshitz_1960}, where the wavefunctions cross over from populating a single bilayer, to being spread over all layers [Fig.~\ref{fig:1}(e)].
This allows to mix the carriers that were formerly separated by their minivalley character.
To summarize, by gating, carriers residing in the minivalleys can be addressed individually, filtered and mixed. 
Such minivalley control makes TDBG with an intermediate angle a promising platform for new efficient type of electronics, called valleytronics.

Here we use high quality TDBG devices to investigate the tunable minivalley polarization and interlayer coupling at a twist angle of 2.37$^\circ$.
Using a dual-gated structure, we first map out the different minivalley states by measuring Shubnikov-de Haas (SdH) oscillations as a function of total density and displacement field. We analyze the measurements using a capacitance model and bandstructure calculations. 
Then, we interpret the transition between coupled and decoupled layers in terms of a Lifshitz transition.

We twist two graphene bilayers using the tear and stack method~\cite{Kim_2016_fab}, and fabricate a multi-terminal device as shown in Fig.~\ref{fig:1}(c) (for details see Supplemental Material~\cite{SM}).
The TDBG stack is sandwiched between two layers of hexagonal boron nitride (hBN). 
On one side a graphite layer, used as a global bottom gate $V_\mathrm{bg}$, is added.
The entire stack is transferred onto a Si/SiO$_2$ (285~nm) substrate. 
We define the TDBG mesa by reactive ion etching, and evaporate Cr/Au to form the Ohmic edge contacts [yellow in Fig.~\ref{fig:1}(c)].
Then two layers of top gates, separated by an aluminum-oxide dielectric, are deposited on top of the stack.
Throughout this manuscript we bias the two top gates such that they together act as a single global top gate $V_\mathrm{tg}$.
All measurements are performed in a dilution cryostat at a temperature of 70~mK, unless stated otherwise.
We use a two-terminal voltage bias setup to obtain the conductance $G$, or a four-terminal current bias setup to measure the longitudinal and Hall resistances, $R_\mathrm{xx}$ and $R_\mathrm{xy}$, both with standard lock-in techniques (see Supplemental Material~\cite{SM}). The magnetic field is applied perpendicular to the plane of the sample.
The sample contains three similar devices, one of which is used in the main text, the other two, as well as other contact configuration of the first, are presented in the Supplemental Material~\cite{SM}.

\begin{figure*}[t]
\includegraphics{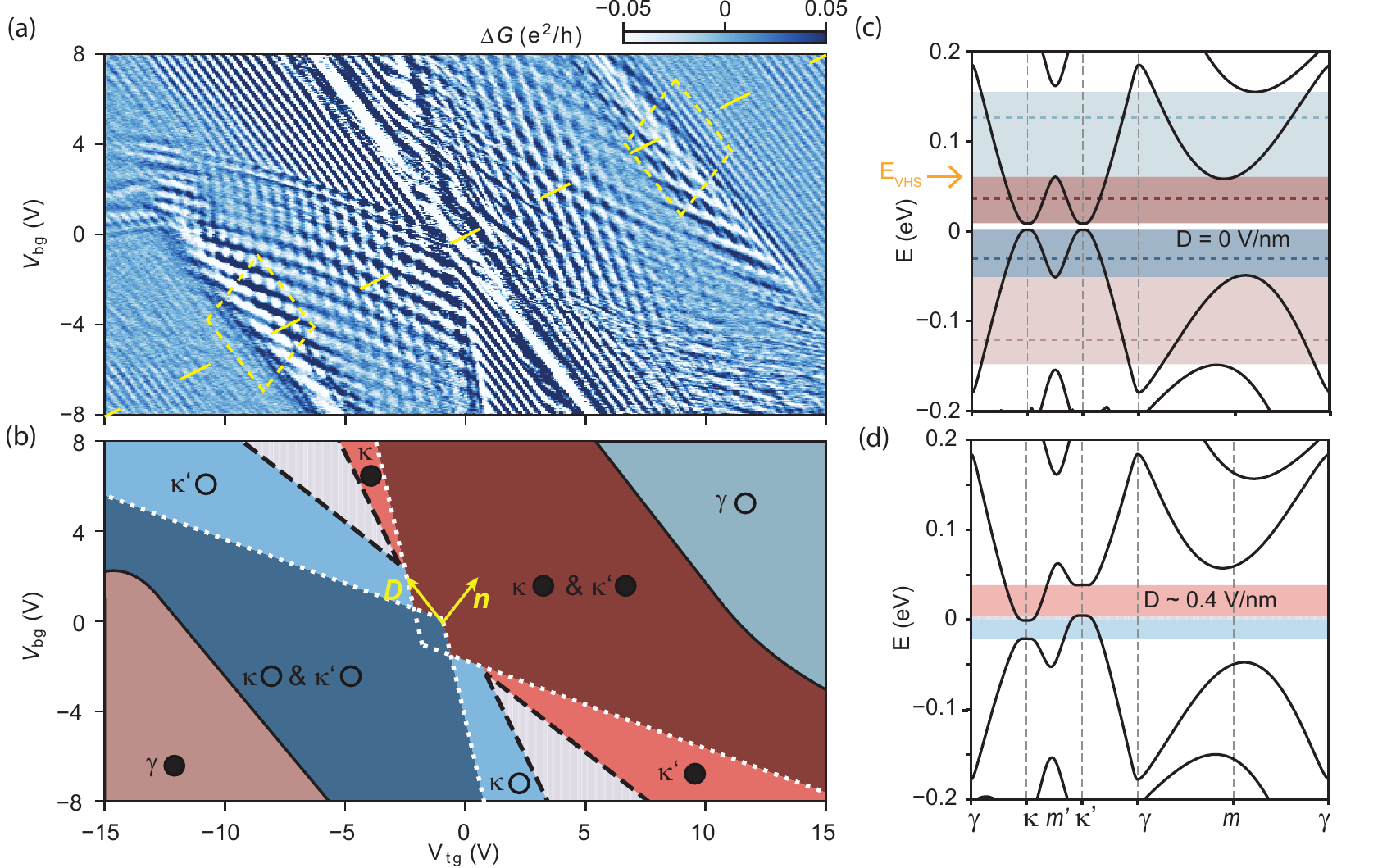}
\caption{\label{fig:2} (a) The conductance modulation $\Delta G$, measured between contacts 1\,\&\,2 and 3\,\&\,4 [Fig.~\ref{fig:1}(d)] shows Shubnikov-de Haas oscillations as a function of top gate $V_\mathrm{tg}$ and bottom gate $V_\mathrm{bg}$, at a magnetic field $B=2$~T. 
Gate voltages used in Fig.~\ref{fig:3}(a,b-c) are plotted with the dashed yellow line along the diagonal, and the dashed yellow boxes, respectively.
(b) Phase diagram of the carriers present as a function of $V_\mathrm{tg}$ and $V_\mathrm{bg}$. The dotted white and dashed black lines are calculated using the capacitor model discussed in Ref.~\cite{Rickhaus_2020_CDW} and the solid black lines are sketched. The regions are colored according to the electronic states being occupied (red) or empty (blue), and the characteristic point from which the Fermi surfaces originate. The directions of the total density $n$ and displacement field $D$ are displayed by the yellow arrows.
(c,d) Bandstructure calculations at zero $D$ (c) and $D\sim\,0.4$~V/nm (d). The color of the highlighted energy windows correspond to the regions indicated in (b). The energy at which a van Hove singularity occurs $E_\mathrm{VHS}$ at the $m'$-point is highlighted by an orange arrow. See the Supplemental Material for details on the calculations~\cite{SM}.}
\end{figure*}

To study the phase diagram of our TDBG device we present the conductance modulation $\Delta G$, after subtracting a smooth background from the measured conductance $G$, as a function of $V_\mathrm{tg}$ and $V_\mathrm{bg}$ at a magnetic field $B=2$~T in Fig.~\ref{fig:2}(a). 
Graphs of the conductance at various $B$ and details on the background subtraction are presented in the Supplemental Material~\cite{SM}, as well as an estimation of the carrier mobility of $\sim$~25.000~cm$^2$/V$\cdot$s and a mean free path of $\sim$~350~nm.
For clarity we depicted the origin and directions [yellow arrows in Fig.~\ref{fig:2}(b)] of the total density $n\propto C_\mathrm{tg}V_\mathrm{tg}+C_\mathrm{bg}V_\mathrm{bg}$ and displacement field $D\propto C_\mathrm{bg}V_\mathrm{bg}-C_\mathrm{tg}V_\mathrm{tg}$, respectively, where $C_\mathrm{i}$ is the capacitance between the TDBG and gate electrode $i$.
The SdH oscillations in Fig.~\ref{fig:2}(a) reveal two types of regions.
First, regions with a single set of SdH oscillations along constant $n$ show up (e.g. around $V_\mathrm{bg},V_\mathrm{tg}=8$~V, $-10$~V), implying the existence of a single carrier gas of density $n$.
Second, we observe regions with two sets of SdH oscillations that cross each other (e.g. around $V_\mathrm{bg},V_\mathrm{tg}=4$~V, $4$~V).
We interpret this as having two decoupled carrier gases occupying the two separate bilayers, one that is coupled more strongly to the top gate and one that is coupled more strongly to the bottom gate.
The overall asymmetry with respect to the $D$-axis is caused by a combination of the inherent electron-hole asymmetry of TDBG~\cite{Lee_2019} and the recently described built-in crystal field~\cite{Rickhaus_2019_cfield}.
We sketch the different regions of the phase diagram in Fig.~\ref{fig:2}(b).
In order to identify on which layers and minivalleys the carriers reside in the different regions we compare the data to a capacitor model and bandstructure calculations.

The results of the capacitor model  are presented as the dotted white and dashed black lines in Fig.~\ref{fig:2}(b) (for details see Supplemental Material~\cite{SM}), which represent the coincidence of the Fermi energy $E_\mathrm{F}$ with band extrema of parabolic electron and hole minibands at $\kappa$ and $\kappa'$.
These lines separate different regions of band population, colored according to the electronic states being occupied (electrons) or empty (holes), red and blue, respectively.
In the dark red/blue regions both $\kappa$ and $\kappa'$ minibands are populated, resembling the regions having two sets of SdH oscillations with different slopes in Fig.~\ref{fig:2}(a). 
The coupling [slopes in Fig.~\ref{fig:2}(a,b)] of the two carrier gases to $V_\mathrm{tg}$ and $V_\mathrm{bg}$ differs because their wavefunctions are mostly localized in separate layers [Fig.~\ref{fig:1}(e)], causing one of the two gates to be screened by the carriers in the other layer.
The light red/blue regions [see also Fig.~\ref{fig:2}(d)] highlight the parts where only one of the two minibands ($\kappa$ or $\kappa'$) is populated due to the finite $D$ that both increases the band gaps $\Delta_\mathrm{D}$ and shifts them with respect to each other. 
Since only the band at $\kappa$ or $\kappa'$ is populated, the device is completely minivalley polarized in these regimes.
This is in accordance to the single set of SdH oscillations parallel to $D$ observed in Fig.~\ref{fig:2}(a). 
There, the Fermi energy $E_\mathrm{F}$ in the other layer resides in the gap, such that no screening occurs.
The grey area enclosed by the dashed black lines represents overlap of electron and hole minibands from $\kappa$ and $\kappa'$, and vice versa. In this region the single particle picture breaks down. Therefore, it is outside the scope of this manuscript, but is treated in detail in Ref.~\cite{Rickhaus_2020_CDW}.
Up to this point, we identified several minivalley polarized (layer decoupled) states, all based on the wavefunction of the carriers at $\kappa$ and $\kappa'$ [Fig.~\ref{fig:1}(e)]. 
Electrons or holes are present in either of the two minivalleys separately, both minivalleys can be populated with the same carrier type, or they can be populated with different carrier types.
We thus have full control over the minivalley polarization of the bulk charge transport in our TDBG device.

When tuning from charge neutrality to larger $E_\mathrm{F}$ (and $n$) we expect to enter a regime where the carriers at $E_\mathrm{F}$ encircle the $\gamma$-point [Fig.~\ref{fig:1}(d)].
The wavefunction of the $\gamma$-carriers is spread over all four layers as sketched in Fig.~\ref{fig:1}(e) and therefore the two bilayers are coupled.
Indeed, we do observe a single set of SdH oscillations in Fig.~\ref{fig:2}(a) above a certain density threshold. 
Therefore, we hypothesize that a Lifshitz transition, a change of the Fermi surface topology, occurs along the solid black line in Fig.~\ref{fig:2}(b). Such a Lifshitz transition has been observed before in TBG~\cite{Kim_2016_lifshitz,Cao_2016}. We extend the validity and range of these pioneering works by adding independent control over displacement field and density.

For further understanding, the bandstructure is calculated using a continuum model, where linear potentials between the layers are used to model the displacement field. A crystal field of 0.1~V/nm is included~\cite{Rickhaus_2019_cfield}, leading to a total gap at zero $D$ of $\sim\,$7~meV.
The different carrier regions discussed before are apparent in the bandstructures calculated at $D=0$~V/nm and $D\sim\,0.4$~V/nm in Fig.~\ref{fig:2}(c,d).
We observe that the parabolic band picture is limited to a Fermi energy $E_\mathrm{F}$ of at most half of the energy of the van Hove singularity $E_\mathrm{VHS}$ at the $m'$- and $m$-point~\cite{Li_2010}.
Also, at $E_\mathrm{F}=E_\mathrm{VHS}$ the center(s) of the Fermi surface(s) changes from the $\kappa$- and $\kappa'$-points to the $\gamma$-point. The Lifshitz transition should thus involve a change in both between electrons and holes, and degeneracy.

\begin{figure}[t!]
\includegraphics{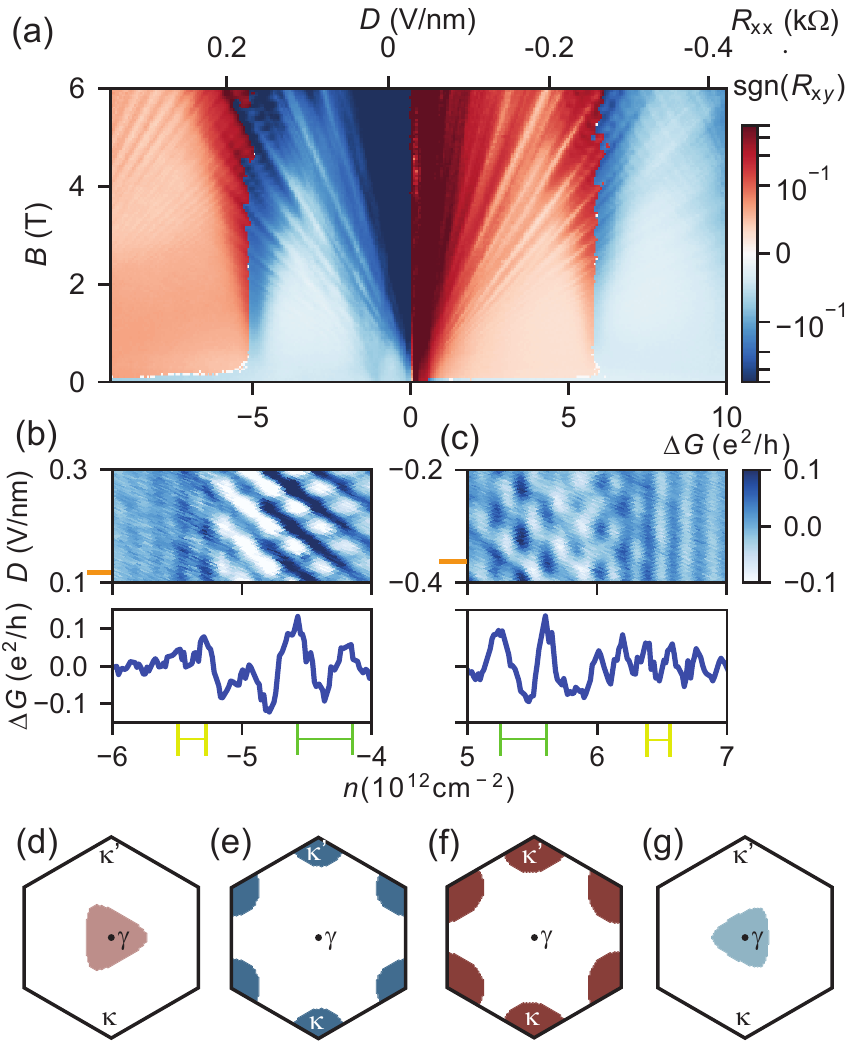}
\caption{\label{fig:3} (a) Longitudinal resistance $R_\mathrm{xx}$ measured between contacts 2 and 4 times the sign of the Hall resistance $R_\mathrm{xy}$ measured between contacts 3 and 4 with a current applied from contact 1 to 6, as a function of density $n$ and displacement field $D$, and magnetic field $B$. A symmetric logarithmic color-scale is used, where the section around zero is linearly interpolated. 
(b,c) $\Delta G$ as a function of $D$ and $n$, including a linetrace obtained by binning the data along the $D$-axis, at a value indicated by the orange ticks. The change in oscillation period is indicated by the green and yellow spacers.
(d-g) Fermi surfaces drawn in the mini-Brillouin zone calculated at $D=0$~V/nm and energies as indicated in Fig.~\ref{fig:2}(c). The closed surfaces are colored following Fig.~\ref{fig:2}(b).}
\end{figure}

In Fig.~\ref{fig:3}(a) we present the longitudinal resistance $R_\mathrm{xx}$ times the sign of the Hall resistance $R_\mathrm{xy}$ as a function of $B$ while changing $n$ and $D$ at the same time, spanning the diagonal of the gate-gate map in Fig.~\ref{fig:2}(a) (see Supplemental Material for constant $D$ measurements).
The color thus represents the carrier type (red for electrons, blue for holes) and the contrast is given by $R_\mathrm{xx}$.
We observe changes in the sign of $R_\mathrm{xy}$ implying a crossover between electrons and holes, and Landau level fans of the corresponding electron and hole Fermi surfaces. 
The Landau fans emerging at nonzero densities originate from the fully filled first moir\'{e} band.
Apart from the expected change in carrier type at the charge neutrality point ($n=0$) we observe two more sign changes, at $n=$~\SI{-5.25e12}~cm$^{-2}$ and $n=$~\SI{5.6e12}~cm$^{-2}$.
At these densities the electron and hole Landau fans emerging from different points ($\kappa$ and $\kappa'$, and $\gamma$, respectively) cross.
Using the densities where the Landau fans emerge and where they cross (corresponding to $E_\mathrm{VHS}$) we estimate the twist angle of 2.3$^\circ$, which we are able to refine with the Hofstadter butterfly spectrum to 2.37$^\circ$, see Supplemental Material~\cite{SM}. 
Additionally, at the point where $R_\mathrm{xy}$ changes sign, the degeneracy changes by a factor 2 for a trace at constant $D$. The periodicity of the SdH oscillations as a function of $n$ [indicated by the yellow and green spacers in Fig.~\ref{fig:3}(b,c)] provides us with $\delta n=$~\SI{1.9e11}~cm$^{-2}$ and $\delta n=$~\SI{3.9e11}~cm$^{-2}$. This change in periodicity of the Landau levels as a function of density corresponds to a change in degeneracy from $\sim\,$4 to $\sim\,$8, respectively, where a degeneracy of 4 stems from the spin and valley degrees of freedom in bilayer graphene, and the factor 2 stems from the layer degeneracy.
We thus confirm the Lifshitz transition, where the Fermi surface changes from two electron/hole surfaces around $\kappa$ and $\kappa'$ to a single hole/electron Fermi surface around $\gamma$. We illustrate this by presenting the calculated Fermi surfaces for $D=0$ in Fig.~\ref{fig:3}(d-g).
We can thus either populate states where the two bilayers are decoupled and individually controllable, or where they are coupled, leading to a single four-layer system [Fig.~\ref{fig:1}(e)].
Therefore, we can not only address the minivalleys individually but also mix the corresponding wavefunctions by passing through the Lifshitz transition. In addition, locally tuning across the Lifshitz transition can potentially be exploited to control the valley degree of freedom~\cite{DeBeule_2019}.

In conclusion, we have used dual-gated TDBG with an intermediate twist angle to demonstrate control over the minivalley polarization and the interlayer coupling.
Using local gates to select $\kappa$-, $\kappa'$-character in the bulk charge transport enables future engineering of valley valves and filters~\cite{DeBeule_2019}.
Perfect valley polarization in the bulk offers greater device flexibility in comparison to realizations where the valley polarized transport only occurs at natural edges or topological boundaries.
Combining this with the empty states and states with $\gamma$-character, opens the possibility to develop logic gates based on the valley degree freedom~\cite{Ang_2017}.
These potential applications show that TDBG with an intermediate angle is a promising route towards valleytronics~\cite{Rycerz_2007,Schaibley_2016}.\newline

All data used in this Letter is made available online at the ETH Zurich research collection \href {https://doi.org/10.3929/ethz-b-000438922}{DOI 10.3929/ethz-b-000438922}.

\begin{acknowledgements}
We acknowledge the support of the ETH FIRST laboratory, and financial support from the European Graphene Flagship and the Swiss National Science Foundation via NCCR Quantum Science.  
P. Rickhaus acknowledges financial support from the ETH Fellowship program. 
Growth of hexagonal boron nitride crystals was supported by the Elemental Strategy Initiative conducted by MEXT, Japan and the CREST (JPMJCR15F3), JST. 
\end{acknowledgements}

%P.R. and G.Z. fabricated the devices. 
%F.K.d.V., P.R. and E.P performed the measurements. 
%J.Z. performed the bandstructure calculations, under supervision of A.H.McD. 
%T.T and K.W. supplied the hBN crystals. 
%F.K.d.V., J.Z., E.P., M.M., A.K., A.H.McD., K.E., T.I. and P.R. discussed the data. 
%K.E., T.I. and P.R. supervised the project. 
%F.K.d.V. wrote the manuscript with comments from all authors.


\clearpage
\newpage
\beginsupplement
\onecolumngrid

\section*{\large{Supplemental Material}}
\vspace{10mm}

\section{Continuum model bandstructure calculation}
Twisted double bilayer graphene is constructed by twisting two Bernal stacked bilayers with each other. We adopt the effective low-energy continuum model by plane wave expansion for the inner two layers, as in the twisted bilayer graphene case~\cite{Bistritzer_2011_supp}. When choosing the basis
$
\psi^{T}_k = (c_{1A}(k), c_{1B}(k),c_{2A}(k), c_{2B}(k),c_{3A}(k), c_{3B}(k), c_{4A}(k), c_{4B}(k))
$,
where $1 - 4$ are layer indices and $A,B$ are sublattices, the $\mathbf{k}$-dependent Hamiltonian is
\begin{equation}
H(\mathbf{k}) = 
	\begin{pmatrix}
		h_{\mathbf{k}-\mathbf{K_1}}(\theta/2) & t & 0 & 0 \\
		t^{\dagger} & h_{\mathbf{k}-\mathbf{K_2}}(\theta/2) & T & 0 \\
		0 & T^{\dagger} & h_{\mathbf{k}-\mathbf{K_3}}(-\theta/2) & t \\
		0 & 0 & t^{\dagger} & h_{\mathbf{k}-\mathbf{K_4}}(-\theta/2)
	\end{pmatrix}
\end{equation}
where $\mathbf{K}_j$ is the Dirac point of $j$-th layer and for twisted double bilayer graphene, $\mathbf{K}_1 = \mathbf{K}_2$ and $\mathbf{K}_3 = \mathbf{K}_4$. On diagonal blocks,
\begin{equation}
h_{\mathbf{q}}(\theta) = \hbar v_F
	\begin{pmatrix}
		0 & \xi e^{i\xi(\theta-\theta_\mathbf{q})} \\
		\xi e^{-i\xi(\theta-\theta_\mathbf{q})} & 0
	\end{pmatrix}
\end{equation}
is the Dirac Hamiltonian rotated by $\theta$ and $\theta_{\mathbf{q}}$ is the angle of crystal momentum $\mathbf{q}$ measured from the Dirac point, where $\xi=\pm 1$ is the valley index.

The tunneling between graphene layers of Bernal stacked bilayer is denoted by $t$. We take
\begin{equation}
t=
	\begin{pmatrix}
	0 & 0 \\
	\gamma_1 & 0
	\end{pmatrix}
\end{equation}
where $\gamma_1=330$ meV. The tunneling, $T(\mathbf{r}) = w\sum\limits_{j=1}^3 e^{-i\mathbf{Q}_{j} \cdot \mathbf{r}}T_j$, between the inner two graphene layers is captured by three transfer momenta $\mathbf{Q}_j$ and their corresponding tunneling matrices $T_j$:

\begin{equation}
\mathbf{Q}_1 = (0, -1)\theta k_D, \ \ \mathbf{Q}_2 = (\frac{\sqrt{3}}{2}, \frac{1}{2})\theta k_D,  \ \ \mathbf{Q}_3 = (-\frac{\sqrt{3}}{2}, \frac{1}{2})\theta k_D
\end{equation}

\begin{equation}
T_1 = 
\begin{pmatrix}
1 & 1 \\
1 & 1
\end{pmatrix}, \ \ 
T_2 = 
\begin{pmatrix}
e^{i2\pi/3} & 1 \\
e^{-i2\pi/3} & e^{i2\pi/3}
\end{pmatrix}, \ \ 
T_3 = 
\begin{pmatrix}
e^{-i2\pi/3} & 1 \\
e^{i2\pi/3} & e^{-i2\pi/3}
\end{pmatrix}
\end{equation}

where $k_D=4\pi/3a_0$ with $a_0$ the graphene lattice constant. With good accuracy, the Hamiltonian in our model is expanded to a cut-off moir\'e reciprocal lattice vector $\mathbf{G}_c$.\\
\\
The different sign of the electron-hole asymmetry with respect to the data is a result of the strength of the crystal field and the inherent electron-hole asymmetry that are both input parameters for the continuum model. Both can differ when taking into account the lattice deformation for example, as further discussed in Ref.~\cite{Rickhaus_2020_CDW_supp}.
This discrepancy, however, does not change the interpretation of the data.

\begin{figure}[h]
\includegraphics{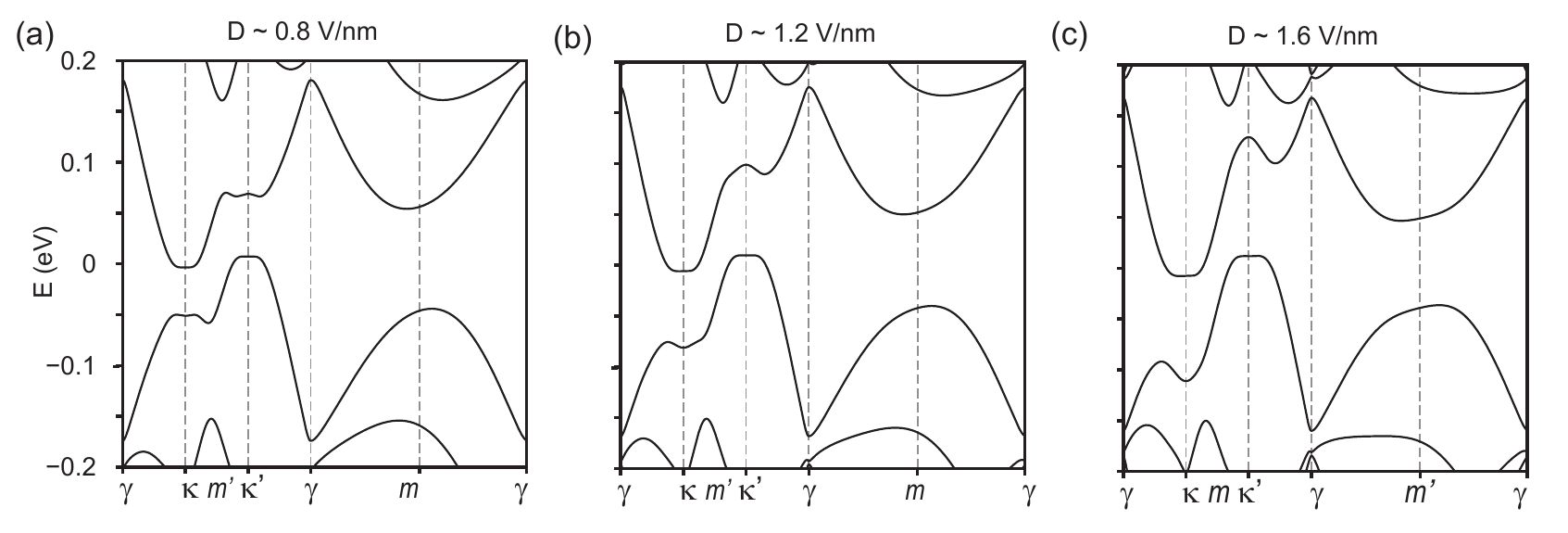}
\caption{\label{fig:S1} Bandstructure calculations at the indicated displacement fields $D$. As $D$ is increased the occupation of $\kappa$ and $\kappa'$ changes up to the point where in one of the two minivalleys the carrier type switches.}
\end{figure}

\begin{figure}[h]
\includegraphics{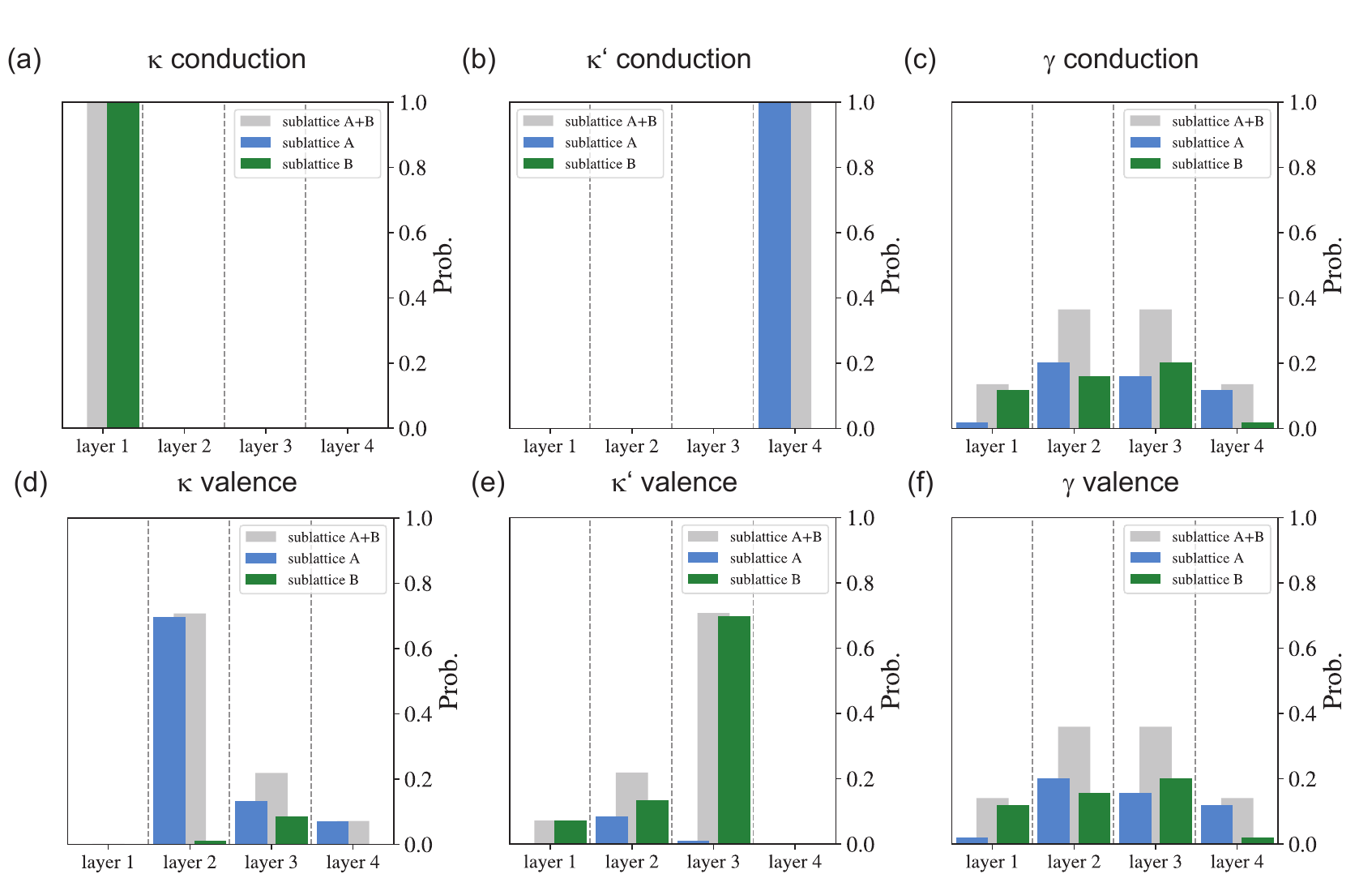}
\caption{\label{fig:S2} Calculated probability distributions of the carrier wave functions residing in the conduction (a-c) and valence (d-f) minibands, for the characteristic points in the bandstructure $\kappa$, $\kappa'$ and $\gamma$. The probabilities are given for both the layers of the tetra-layer stack and for the two sublattices A and B of each of these layers. The total occupation per layer is used for the schematic in the main text (sublattice A+B).}
\end{figure}

\clearpage
\newpage

\section{Fabrication}
Our fabrication follows largely Ref.\cite{Kim2016} and is separated in two phases. First, the heterostack is assembled, then contacts and gates are fabricated.

Both hexagonal boron nitride (hBN) and graphene are mechanically exfoliated with scotch tape and deposited on Si/SiO$_2$ (\SI{285}{nm}) wafers.
For the heterostack we use a PDMS stamp with a PC film on top and start by picking up an hBN flake with a thickness of \SI{27}{nm}. For the twisted double bilayer we start by identifying a single Bernal stacked graphene bilayer. Using the top hBN, we pick up half of the bilayer by tearing the bilayer in two pieces, the other half is left on the chip. Then we rotate our substrate with respect to the stack, and pick up the remaining part. Subsequently the bottom hBN (thickness \SI{50}{nm}) and a large piece of graphite are picked up. Finally, the stack is deposited on a pre-patterned marker chip for further processing. 
 
\begin{figure}[h]
\includegraphics[width=13cm]{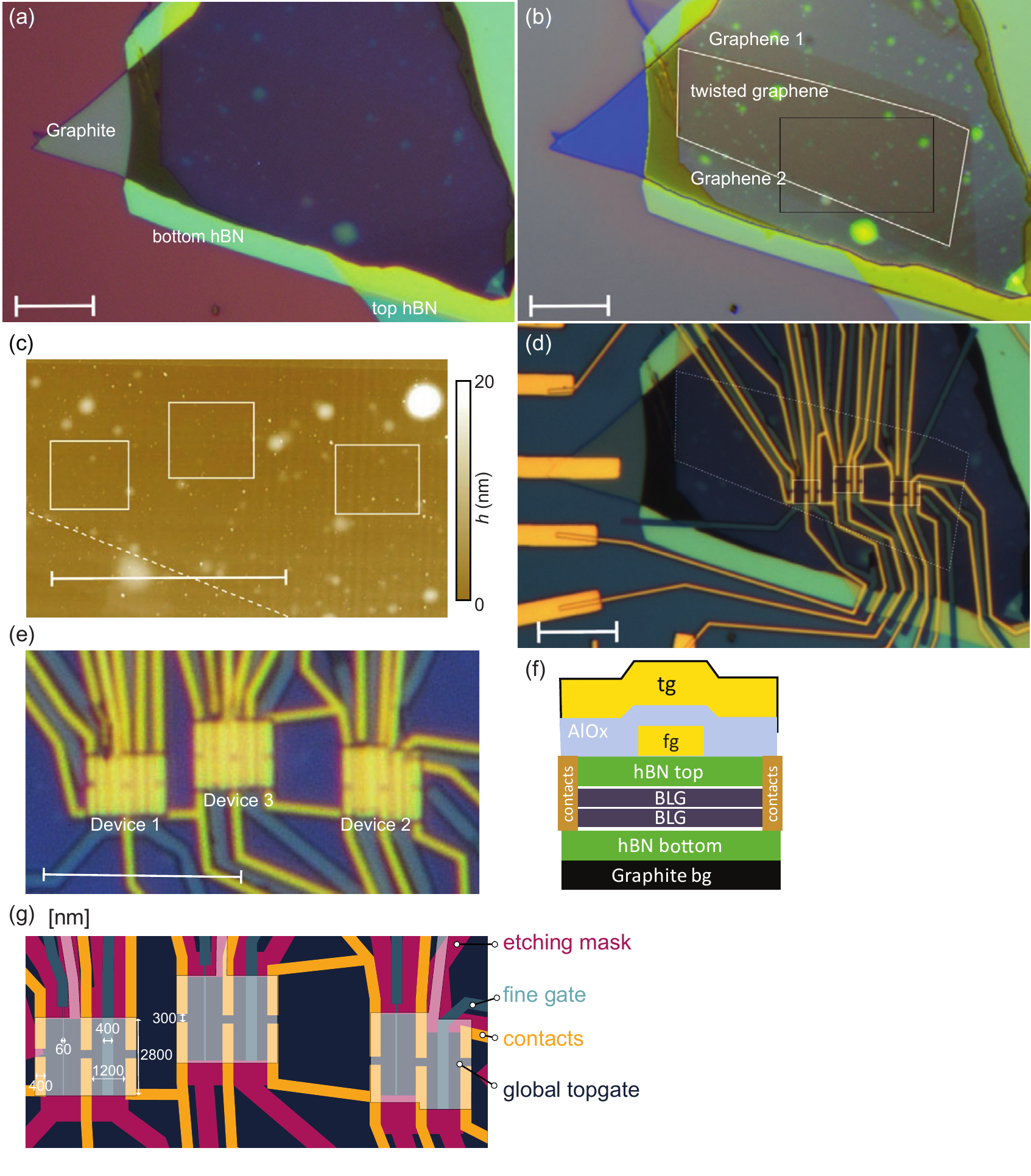}
\caption{\label{fig:S3} (a) Optical microscope image of the stack. Scale-bar for all subfigures: $\SI{10}{\mu m}$.
    (b) Optical microscope image with enhanced contrast. The twisted graphene area is marked with a white line.
    (c) Atomic force microscope (AFM) image of the area which is labelled with a black rectangle in (b). Device areas are indicated with white rectangles.
    (d) Optical microscope image after deposition of contacts, fine gates and after etching.
    (e) Optical microscope image after deposition of the global topgate.
    (f) Schematic cross section of the device, with graphite bottom gate (bg), top and bottom hexagonal boron nitride (hBN), two graphene bilayers (BLG), fine gate (fg), aluminum-oxide (AlO$_\mathrm{x}$), global top gate (tg), and Ohmic contacts sketched.
    (g) Top-view of the design with relevant length scales in nm.
    }
\end{figure}

A picture of the stack in presented in Fig.~\ref{fig:S3}(a). By adjusting the contrast, Fig.~\ref{fig:S3}(b), we identify clean areas on which we perform atomic force microscopy (AFM) measurements, Fig.~\ref{fig:S3}(c). The results are used in order to design the devices on clean and bubble-free areas. 

Afterwards, the device is processed in a clean-room. Cr/Au (10/\SI{50}{nm}) contacts are evaporated after etching the top-hBN layer using a reactive ion etcher (CHF$_3$/O$_2$, $40/4$ sccm, $\SI{60}{W}$). We then fabricate the fine gates using a short plasma (5s, CHF$_3$/O$_2$, $40/4$ sccm, $\SI{30}{W}$) and evaporating  Cr/Au (10/$\SI{70}{nm}$). As a next step, the twisted double bilayer graphene is etched  (CHF$_3$/O$_2$, $40/4$ sccm, $\SI{60}{W}$) and we deposit $\SI{30}{nm}$ of Aluminium Oxide using atomic layer deposition. In a final step, the global topgates are evaporated Cr/Au (10/\SI{70}{nm}).

The relevant length scales of the device are depicted in Fig.~\ref{fig:S3}(g).

\clearpage
\newpage

\section{Measurement setup}
\begin{figure}[h]
\includegraphics{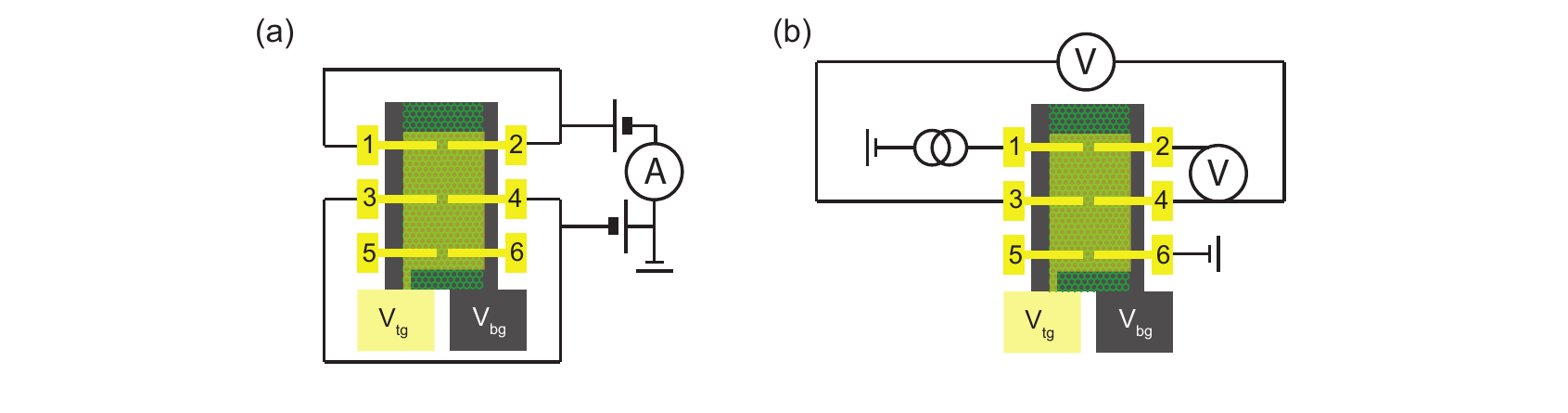}
\caption{\label{fig:S4} (a) Two terminal conductance measurement setup. The voltage is applied symetrically over the source and drain contact and the current is measured. (b) Four terminal current bias setup, where the longitudinal and Hall voltage are measured.}
\end{figure}

For the lock-in measurements presented we make use of two different device configuration, and apply an AC excitation with a typical frequency of 177~Hz. The first is a two terminal, voltage bias, current measure configuration as sketched in Fig.\ref{fig:S4}(a). There we typically use and excitation voltage of 50~uV, and amplify the current 1M times with a current to voltage converter.  For the four terminal, current bias, voltage measure configuration [Fig.\ref{fig:S4}(b)] we typically use 100~nA and use a 1000 times voltage amplifier. 
The conductance and resistance is then calculated straightforwardly from the applied/measured current and voltage, respectively.

\section{Capacitor model}
The capacitor model is an electrostatic description of the multi-layer system where the device stack is assumed to be a series of parallel plate capacitors, influenced by the voltages applied to $V_\mathrm{tg}$ and $V_\mathrm{bg}$. 
The electron and hole minibands at the $\kappa$- and $\kappa'$-point are modeled by their quantum capacitance $C_\mathrm{q}\propto m_\mathrm{eff}$ in the effective mass approximation, with $m_\mathrm{eff}=0.06 m_\mathrm{e}$~\cite{Rickhaus_2020_CDW_supp}, with $m_\mathrm{e}$ the free electron mass. 
For each bilayer, the opening of the gap $\Delta_\mathrm{D}$ as a function of $D$ is taken into account with $D=0.195\Delta_\mathrm{D}$~\cite{Rickhaus_2019_cfield_supp}.
Furthermore, the electron-hole asymmetry is added with an offset of both gaps of 8~meV at zero $D$.
For more details on this model see Refs.~\cite{Rickhaus_2019_thickness_supp,Rickhaus_2019_cfield_supp,Rickhaus_2020_CDW_supp}.

\clearpage
\newpage

\section{Background subtraction}
In order to reveal the SdH oscillations in the conductance measured, we subtract a smooth background. We calculate this background by applying a Savitsky Golay moving filter with 31 points and a polynomial degree of 3 along both the $V_\mathrm{tg}$ and $V_\mathrm{bg}$ axis. Taking the difference between the measured conductance and the smoothened one, we obtain the conductance modulation.

\begin{figure}[h]
\includegraphics{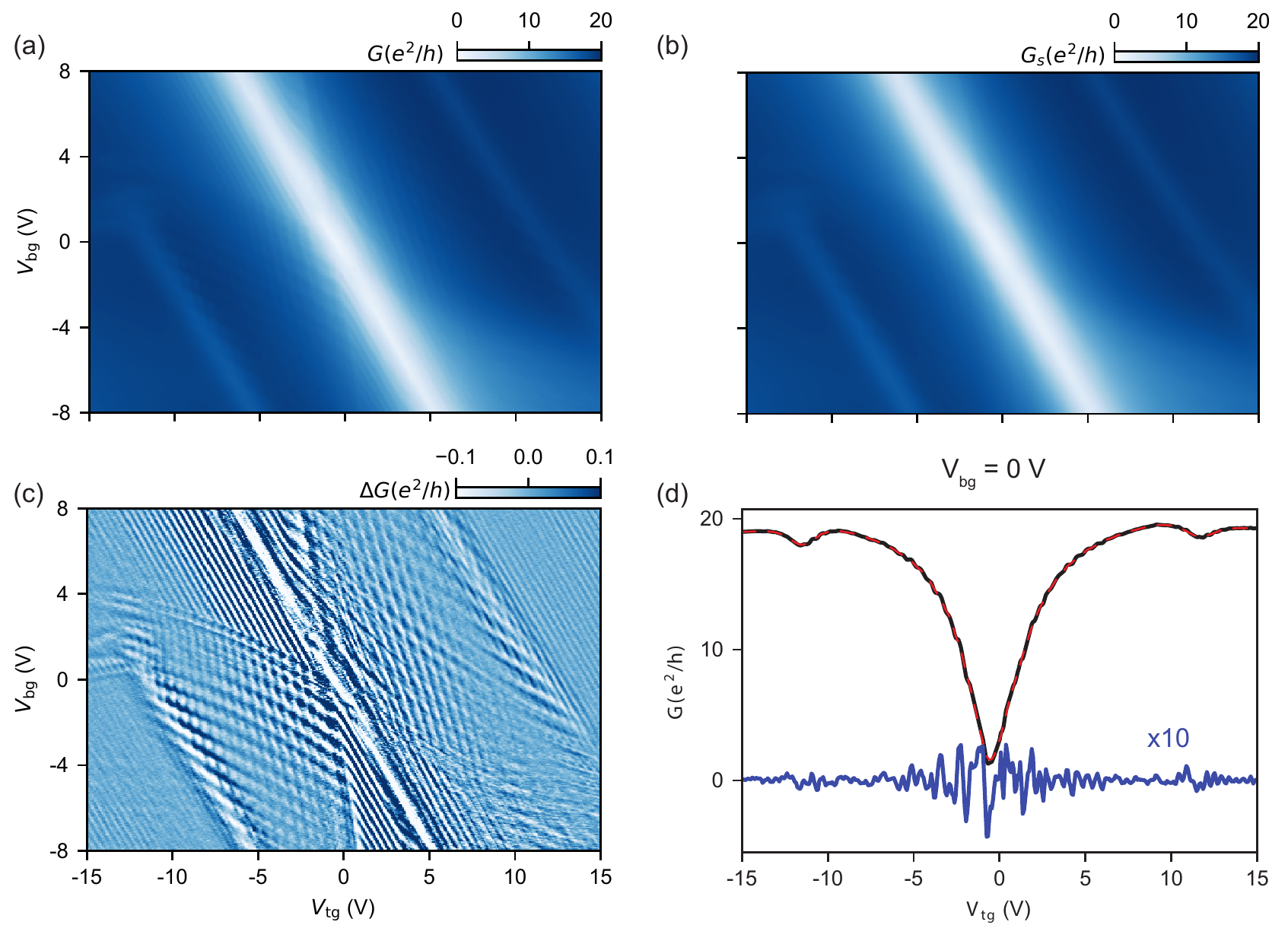}
\caption{\label{fig:S5} (a) Conductance $G$ measured on device 1 between contacts 1\&2 and 3\&4 as a function of $V_\mathrm{tg}$ and $V_\mathrm{bg}$ with a magnetic field of 2~T applied.
(b) Background conductance $G_s$ obtained by smoothing (a).
(c) Conductance modulation $\Delta G$, the result of subtracting (b) from (a).
(d) Example linetrace ($V_\mathrm{bg} = 0$) of the background subtraction procedure. The solid black line is $G$, the dashed red line $G_s$ and the solid blue line $\Delta G$.}
\end{figure}

\clearpage
\newpage

\section{Different contacts device 1}
\begin{figure}[h]
\includegraphics{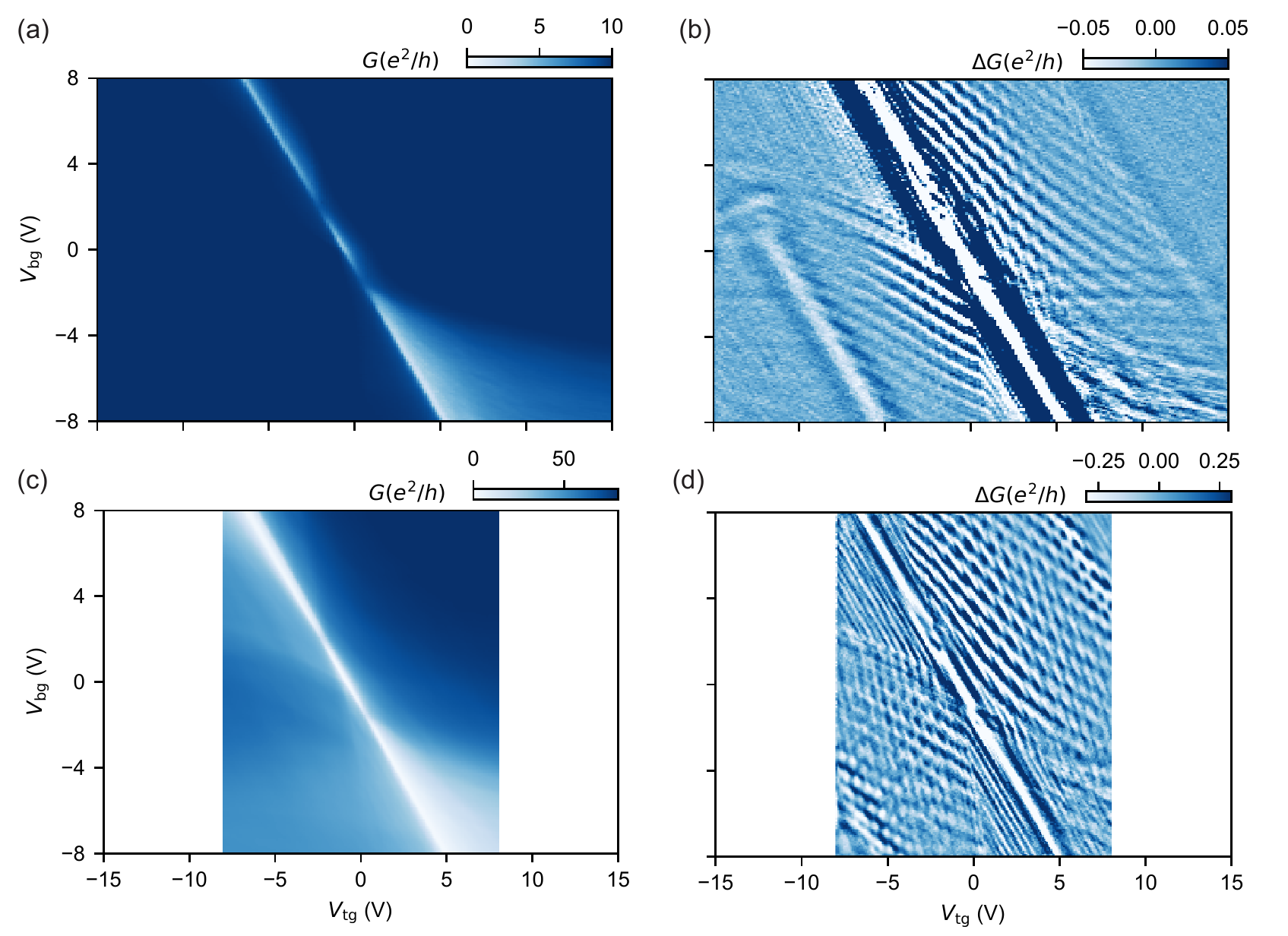}
\caption{\label{fig:S6} (a,c) Conductance $G$ and (b,d)the conductance modulation $\Delta G$ after subtracting a smooth background as a function of $V_\mathrm{tg}$ and $V_\mathrm{bg}$ at $B=2$~T. (a,b) is measured between contacts 3 and 4, and (c,d) between 3\& 4 and 5\& 6. (c,d) is measured at $T=1.5$~K.}
\end{figure}

\clearpage
\newpage

\section{Devices 2 and 3}
\begin{figure}[h]
\includegraphics{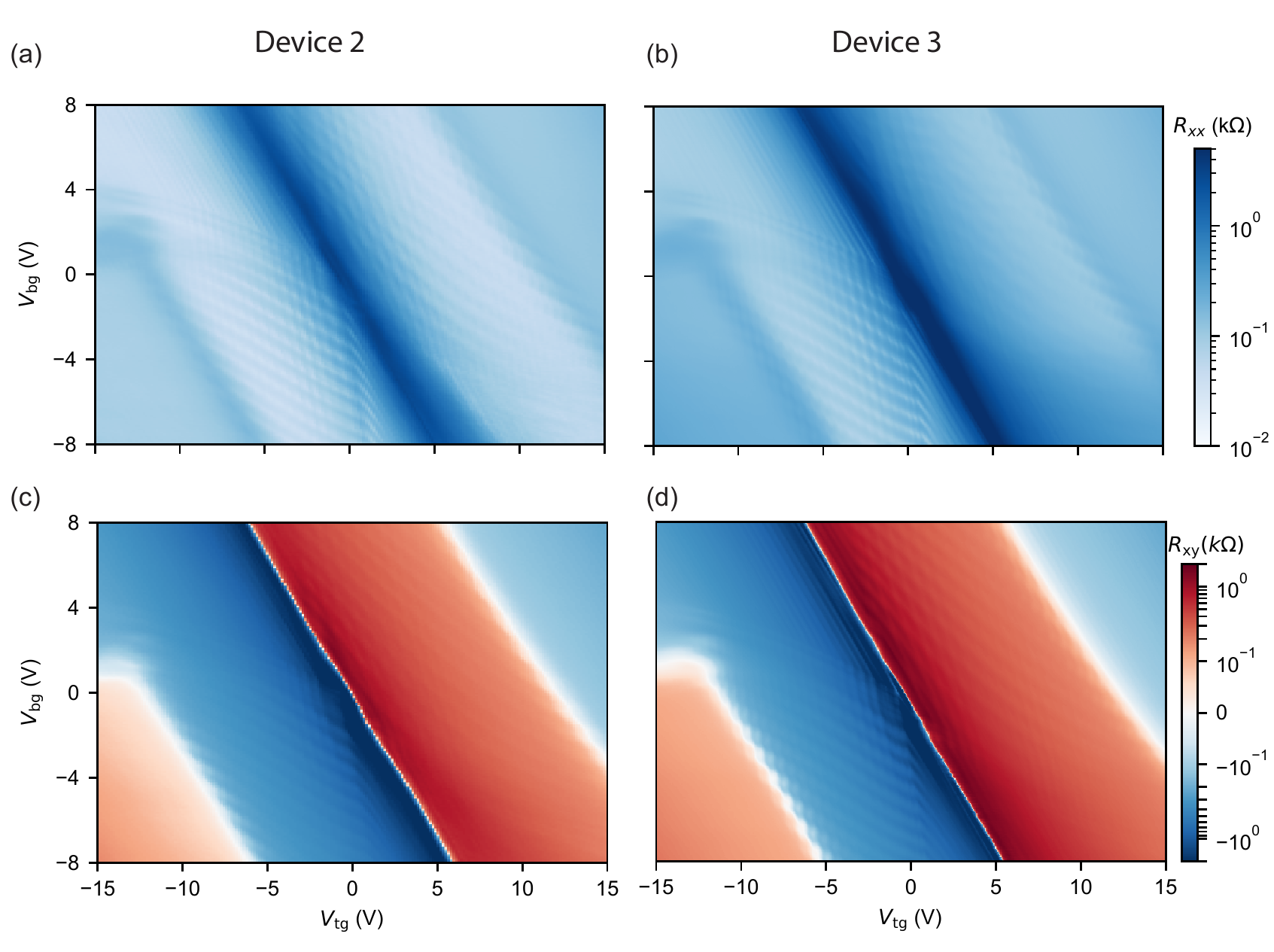}
\caption{\label{fig:S7} (a,c) $R_\mathrm{xx}$ and (b,d) $R_\mathrm{xy}$, both as a function of $V_\mathrm{tg}$ and $V_\mathrm{bg}$ at $B=2$~T measured in device 2 and device 3, respectively. For both $R_{xx}$ is measured between contacts 2 and 4, $R_{xy}$ between 3 and 4, while sending a current from 1 to 6. All four resistance maps show very similar behaviour to device 1 presented in the main text. The different regimes of SdH oscillations are visible, as well as the change of carrier type at the Lifshitz transition. Note that device 3 was measured at 1.7~K.}
\end{figure}

\clearpage
\newpage

\section{Conductance device 1 at 0T and 2T}
\begin{figure}[h]
\includegraphics[width=12cm]{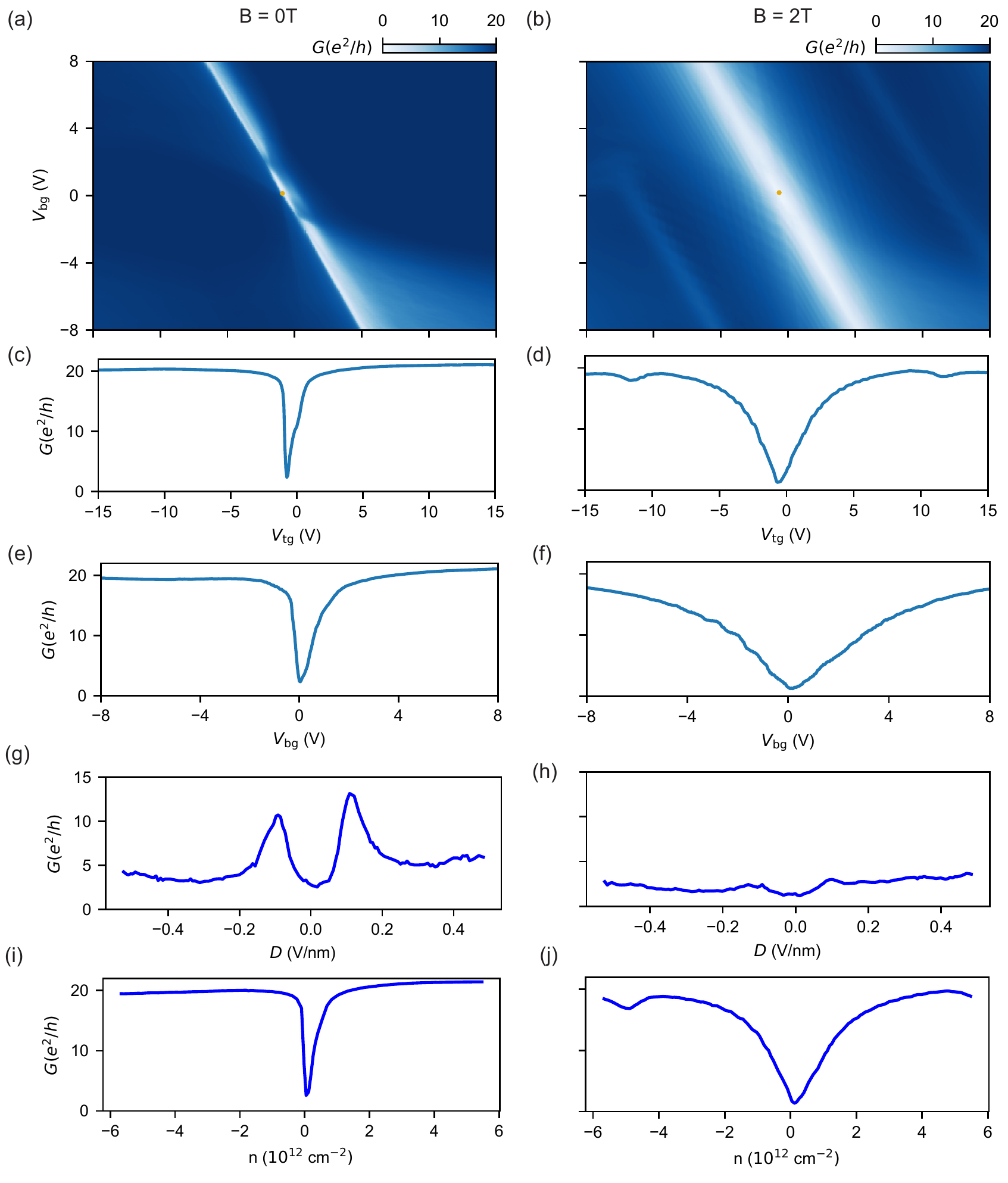}
\caption{\label{fig:S8} (a) Conductance $G$ measured on device 1 between contacts 1\&2 and 3\&4 as a function of $V_\mathrm{tg}$ and $V_\mathrm{bg}$ without a magnetic field applied.
(b) Same as (a) at a magnetic field of 2~T. Both (a,b) have the charge neutrality point indicated by an orange dot.
(c,d) Linetraces at $V_\mathrm{bg} = 0$~V, for $B=0,2$~T, respectively. 
(e,f) Linetraces at $V_\mathrm{tg} = -0.75$~V, for $B=0,2$~T, respectively.
(g,h) Linetraces where the density $n \sim 0$ for $B=0,2$~T, respectively. The points are selected by binning $n$ calculated using $V_\mathrm{tg}$, $C_\mathrm{tg}$, $V_\mathrm{bg}$ and $C_\mathrm{bg}$.
(i,j) Linetraces where the displacement field $D \sim 0$ for $B=0,2$~T, respectively.}
\end{figure}

\clearpage
\newpage

\section{Gate gate maps at 4T and 6T.}
\begin{figure}[h]
\includegraphics[width=15cm]{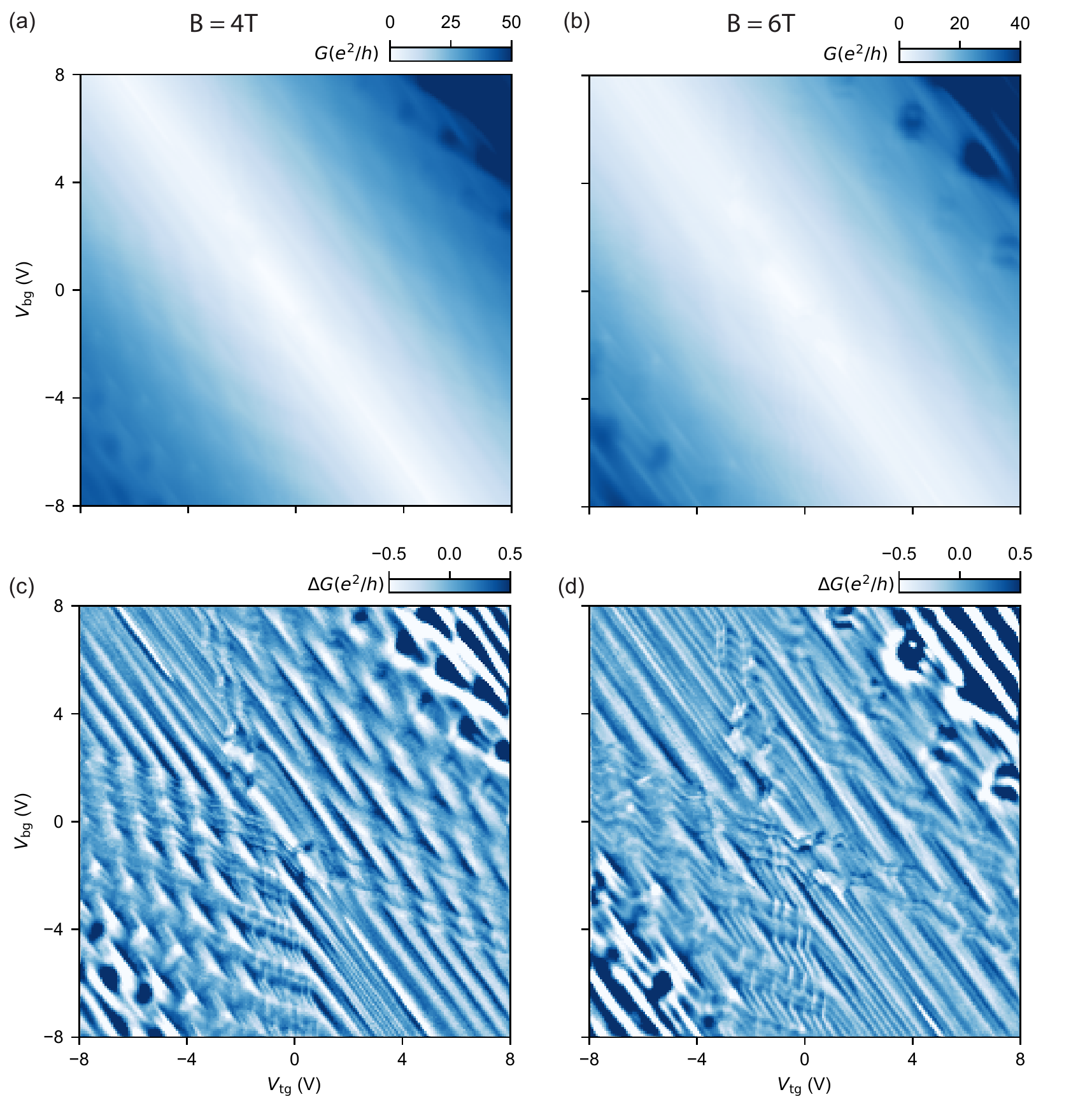}
\caption{\label{fig:S9} (a,b) Conductance $G$ as a function of $V_\mathrm{tg}$ and $V_\mathrm{bg}$ at $B=4$~T and $B=6$~T, respectively. 
(c,d) Conductance modulation $\Delta G$, after subtracting a smooth background, as a function of $V_\mathrm{tg}$ and $V_\mathrm{bg}$ at $B=4$~T and $B=6$~T, respectively. 
All measurements are taken on device 3 from contacts 1\&2 to 3\&4 at a temperature of 1.7~K.}
\end{figure}

\clearpage
\newpage

\section{Mobility and mean free path}
\begin{figure}[h]
\includegraphics[width=14cm]{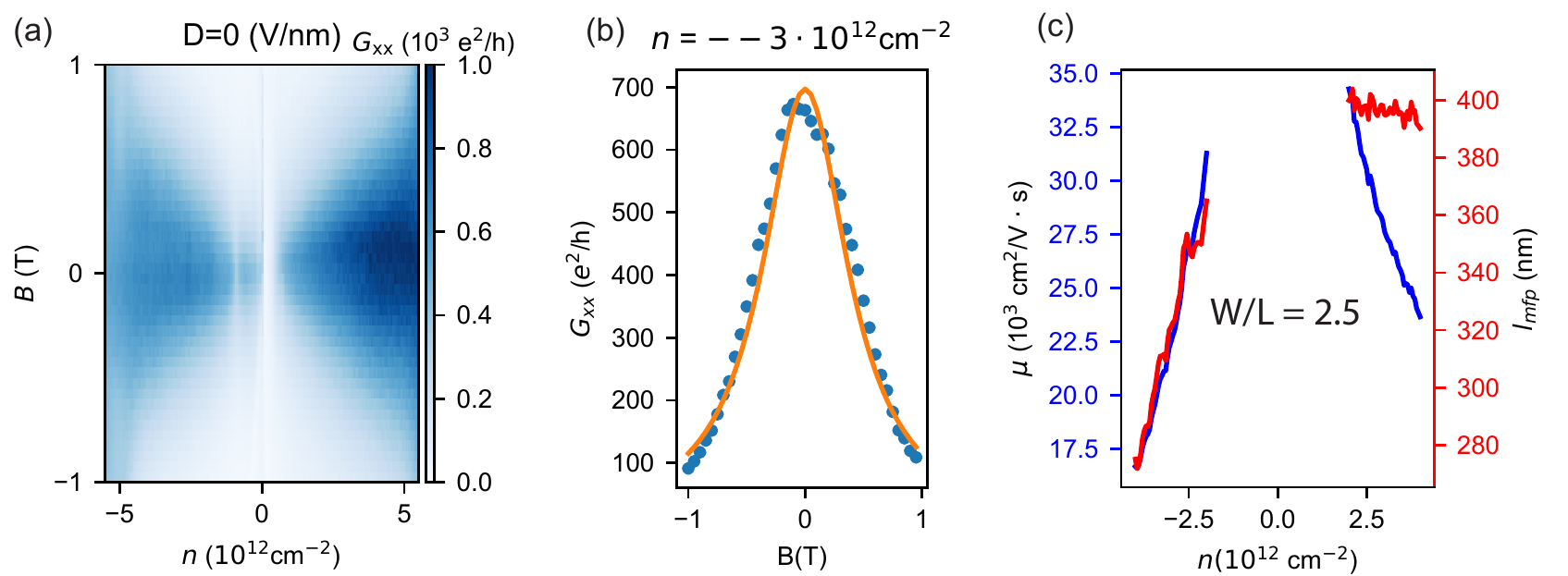}
\caption{\label{fig:S10} (a) Longitudinal conductance $G_{xx}$ as a function density $n$ and magnetic field $B$. 
(b) Example of a fitted linetrace using the classical Drude model at the indicated density.
(c) Carrier mobility $\mu$ and mean free path $l_{mfp}$ as a function of $n$.}
\end{figure}

We estimate the mobility and mean free path of the carriers in our TDBG at zero displacement field. First, we obtain the classical longitudinal magnetoconductance $G_{xx}$, from $R_{xx}$ and $R_{xy}$ via tensor inversion, the result is plotted in Fig.~\ref{fig:S10}(a) ($D=0$). Then we fit this with the classical Drude model for magnetoconductivity:
\begin{equation}
G_{xx}=A\frac{n \mu e}{1+\mu^2 B^2} 
\end{equation}
where $A$ is the geometry factor ($W/L$), $n$ is the carrier density, $\mu$ the carrier mobility, $e$ is the elementary charge and $B$ the magnetic field.
Since our device does not have a proper hall bar shape, we estimated $A$=2.5, first based on the geometry (Fig.~\ref{fig:S1}), and then checked this by making $A$ a fit parameter (next to $\mu$). We disregard all $n$ where the fitted $A$ is more than 20$\%$ different than the estimated value. Finally, we set $A$=2.5 for these densities and fit $\mu$, of which the result is plotted in (Fig.~\ref{fig:S10}(b,c)). 
In addition, we calculate the mean free path following:
\begin{equation}
l_{mfp}=\sqrt{\pi n /2} \frac{\hbar \mu}{e}    
\end{equation}
where a degeneracy of 8 (minivalley, valley, spin) is used and $\hbar$ is the reduced Planck's constant. The mean free path is plotted in Fig.~\ref{fig:S10}(c) as well.

\clearpage
\newpage

\section{Landau fans at constant D}
\begin{figure}[h]
\includegraphics[width=14cm]{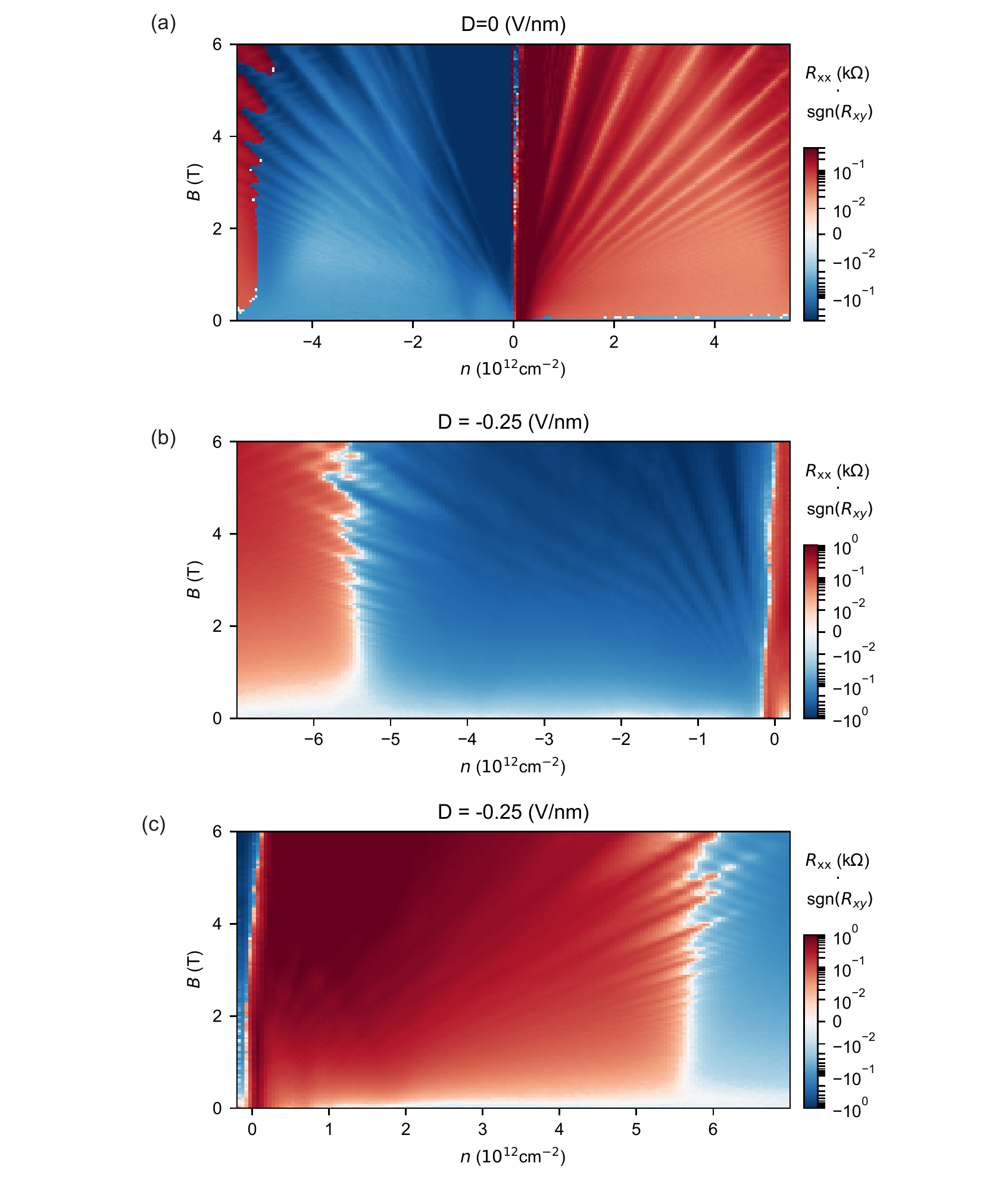}
\caption{\label{fig:S11} (a) Longitudinal resistance $R_{xx}$ measured on device 1 between contacts 2 and 4 as a function of density $n$ and magnetic field $B$, multiplied by the sign of the Hall resistance $R_{xy}$, measured between contacts 3 and 4, for zero applied displacement field $D$, (b) for $D=-0.25$~V/nm and (c) for $D=0.25$~V/nm.}
\end{figure}

\clearpage
\newpage

\section{Angle extraction}
\begin{figure}[h]
\includegraphics[width=14cm]{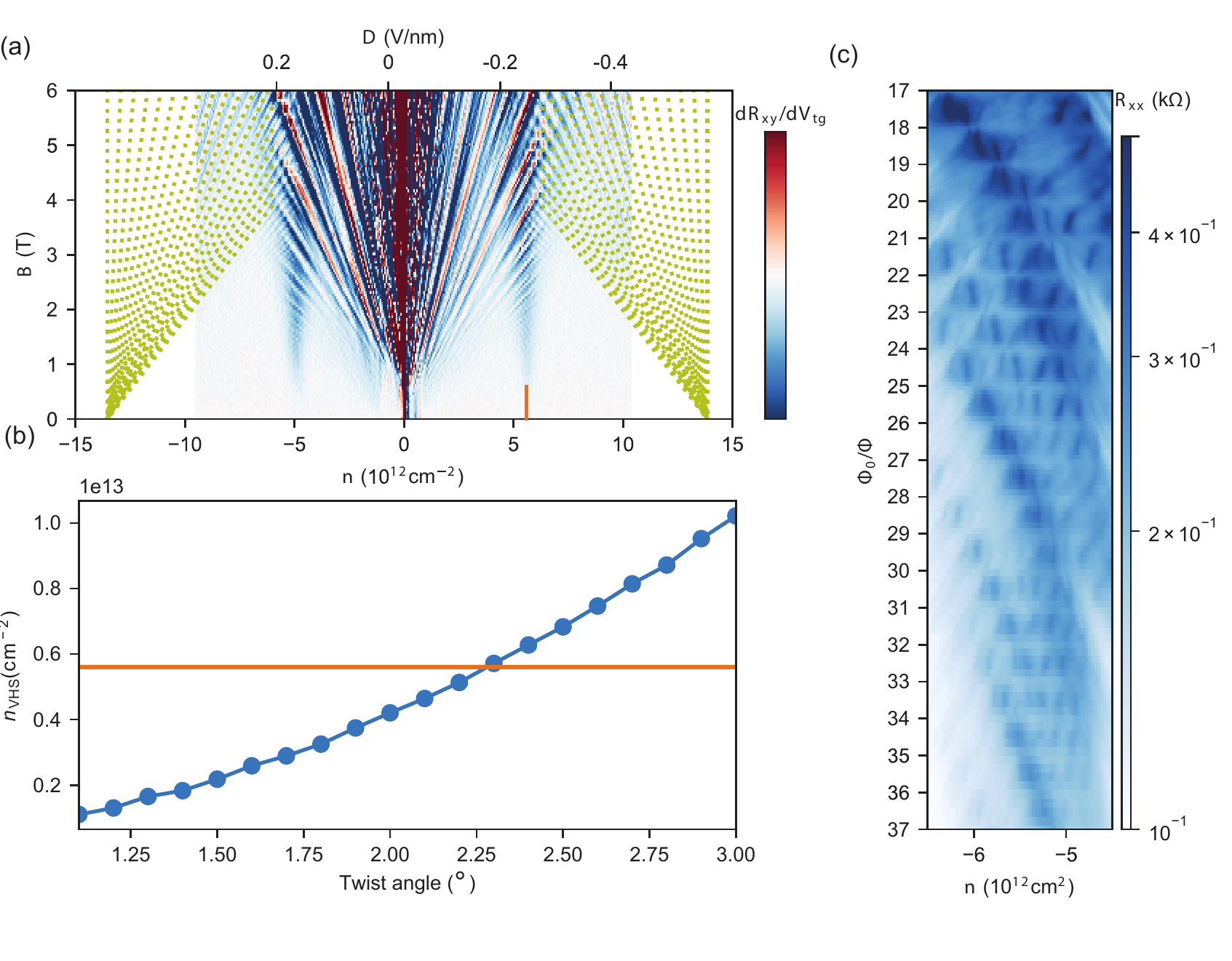}
\caption{\label{fig:S12} We present three ways of estimating the angle of our TDBG. 
(a) The derivative of $R_\mathrm{xy}$ with respect to $V_\mathrm{tg}$ as a function of $n$, $D$ and $B$, measured in device 1 is shown. The Landau fans emerging from full filling of the first miniband are highlighted with the dotted yellow lines. Based on the densities of $n=1.35\cdot 10^{13}$~cm$^{-2}$ found, we estimate the angle to be 2.4$^\circ$. 
(b) The density at which the VHS occurs in the electron miniband $n_\mathrm{VHS}$ is calculated with a continuum model for different twist angles (blue). The orange line represents the density at which we observe the VHS (and Lifshitz transition) of $n=5.6\cdot 10^{12}$~cm$^{-2}$, leading to an estimated angle of 2.3$^\circ$. This angle is used in all bandstructure calculations.
(c) $R_\mathrm{xx}$ as a function of $n$ plotted versus a flux quantum $\Phi_0$ divided by flux $\Phi$ through a single Moire unit cell ($A = 30.5$~nm$^2$), measured in device 2. The periodicity of the horizontal resonances let us extract the twist angle of 2.374$^\circ$ ($\pm 0.002^\circ$).}
\end{figure}

\clearpage

\end{document}